\documentclass[12pt]{article}
\usepackage{amssymb}
\usepackage{amsmath}

\begin{document}

\begin{center}
\vspace{1cm}{\Large {\bf Little Groups of Preon Branes}}

\vspace{1cm} {\bf H. Mkrtchyan} \footnote{ E-mail:hike@r.am} and
{\bf R. Mkrtchyan} \footnote{ E-mail: mrl@r.am} \vspace{1cm}

\vspace{1cm}

{\it Theoretical Physics Department,} {\it Yerevan Physics
Institute}

{\it Alikhanian Br. St.2, Yerevan 375036, Armenia}
\end{center}

\vspace{1cm}
\begin{abstract}

Little groups for preon branes (i.e. configurations of branes with
maximal (n-1)/n fraction of survived supersymmetry) for dimensions
d=2,3,...,11 are calculated for all massless, and partially for
massive orbits. For massless orbits little groups are semidirect
product of d-2 translational group $T_{d-2}$ on a subgroup of
(SO(d-2) $\times$ R-invariance) group. E.g. at d=9 the subgroup is
exceptional $G_2$ group. It is also argued, that 11d Majorana
spinor invariants, which distinguish orbits, are actually
invariant under d=2+10 Lorentz group. Possible applications of
these results include construction of field theories in
generalized space-times with brane charges coordinates, different
problems of group's representations decompositions,
spin-statistics issues.

\end{abstract}

\renewcommand{\thefootnote}{\arabic{footnote}}
\setcounter{footnote}0 {\smallskip \pagebreak }

\section{Introduction and Conclusion}
There is a strong interest to the symmetry structure of M and
related theories. The space-time supersymmetry algebra of
M-theory, is (\cite{TowAsk} \cite{Tow}):

\begin{eqnarray}
\left\{ \bar{Q},Q\right\} &=&\Gamma ^{\mu}P_{\nu}+\Gamma
^{\mu\nu}Z_{\mu\nu}+\Gamma ^{\mu\nu\lambda\rho\sigma}
Z_{\mu\nu\lambda\rho\sigma}, \label{0}\\ \mu, \nu,...
&=&0,1,2,..10. \nonumber
\end{eqnarray}
where $Q$ is Majorana 32-component spinor, $Z_{\mu\nu}$ and
$Z_{\mu\nu\lambda\rho\sigma}$ are membrane and five-brane charges.
This algebra is of (super)-Poincar{\'e} type, i.e. its bosonic
(tensorial Poincar{\'e}) subalgebra is a semidirect product of
Lorentz and Abelian subalgebras, the latter is a sum of $P_\mu$,
$Z_{\mu\nu}$ and $Z_{\mu\nu\lambda\rho\sigma}$. There are
evidences, that this algebra can be extended. Already in the first
works on extended supergravities at d=4 hidden non-linearly
realized symmetries have been found \cite{Crem}, which later on
appeared to be connected (\cite{OP},\cite {dWit}) with space-time
symmetry algebra (\ref{0}), and these observations particularly
lead to suggestions on hidden space-time coordinates and
symmetries of M-theory \cite{dWit2}.  In some yet unknown phase of
M-theory with explicit conformal symmetry the latter is assumed
(\cite{Tow2}, earlier references are \cite {Crem}, \cite {van}) to
be an $OSp(1|64)$, ($OSp(1|32), OSp(1|16), ...$ on lower
dimensionalities) supergroup, which includes (\ref{0}) as
sub-superalgebra, just as usual Poincar{\'e} group $SO(1,3)\ltimes
T_4$ (at d=4) is a subgroup of conformal group SO(2,4).
Generalized space-times are suggested \cite{Gun} for a new
formulations of supergravity/superstring/M-theory, with a
particular purpose of making explicit hidden symmetries of these
theories. The new formulation of higher spin theories is making
use of some of these generalized space-times with $OSp(1|2n)$
symmetry \cite{Vas}\cite{Vas2}, and equivalence to theories in a
usual space-time, in the free case is shown. The recently
discovered \cite{West} non-linear realization representation of
maximal supergravities - d=11 N=1, d=10 N=2a and 2b - make use of
the new space-time algebras, most general of which is the newly
introduced \cite{Olive} "very extended" $E_{11}$ algebra
\cite{West2}.

Correspondingly, the irreducible representations (irreps),
particularly unitary irreps of these and connected algebras are
attracting attention in modern literature.There are specially
interesting representations among them, such as singletons of OSp
(which appear in many circumstances), or preon representations of
super-Poincar{\'e}, which correspond to maximally supersymmetric
BPS branes \cite{Band},\cite{Band2}, and others.

All that is making the investigation of properties of space-time
symmetry algebras of type (\ref{0}) interesting and promising,
both for the purpose of clarifying properties of the theory, and
for seeking new formalisms for it. In the present paper we will
continue investigation \cite{Man}, \cite{mrl3} of the construction
of unitary irreps of algebras (\ref{0}) by applying Wigner's
method of little groups (induction from the little group)
\cite{Wig}\cite{BR}. The main focus of present paper will be the
preonic orbits \cite{Band}, \cite{Band2}, i.e. orbits
(configurations of branes) with maximum fraction of surviving
supersymmetry, some other orbits are considered in previous papers
\cite{Man}, \cite{mrl2}. We find a complete list, at dimensions
2,3,..11 of little groups for the massless representations, i.e.
for those preonic configurations, for which corresponding vector
$p_\mu$ is massless, and all continuous invariants are also zero.
All little groups are of the form of semidirect product of compact
group on group of (d-2)-dimensional translations. These results
are new for d=5,6 (for symplectic-Majorana spinors),7,8,9. For
massive case we find complete answers for dimensionalities 5,8,9.
In this case little groups are compact, since they should be a
subgroup of those for massive particles, i.e. of SO(d-1). We also
show that 11d spinor invariants are actually (12=2+10)d invariant.

There are many possible applications of present results. One is
the determination of the brane content of branes \cite{mrl3}, i.e.
decomposition of irreps w.r.t. the different subalgebras. Among
these algebras can be a tensorial Poincar{\'e} with less number of
tensorial generators, or a usual Poincar{\'e}, so one can consider
a spectrum of usual particle representations in a given brane
representation. This approach provides \cite{mrl3} group-theoretic
interpretation of the results of \cite{Vas2}.

Another application can be the decomposition of different
representation of OSp (which is a "conformal" group for tensorial
Poincar{\'e}) w.r.t. the tensorial Poincar{\'e}.

Next, one has to reconsider \cite{mrl2} the spin-statistics
theorem, with the purpose of generalization of connection between
statistics and spins. Now instead of spin we have a rich set of
representations of little groups, particularly those from Tables 1
and 2, and one has to assign Fermi or Bose statistics to each of
representations of little groups. Moreover, the preon case
resemble strongly the usual two-dimensional situation, when there
is no strong difference between fermions and bosons, both
correspond to the same representation of Poincar{\'e}. Similarly,
in the preon representation of super-Poincar{\'e} with tensor
charges ((\ref{0}), e.g.) the non-zero supercharge is neutral
\cite{mrl2} w.r.t. the little group, supermultiplet has only two
members, both in the same representation of tensorial Poincar{\'e}
and one can expect the generalized statistics, similar to those in
the two-dimensional case.

Last, but not least: one of the most interesting applications (and
initial motivation) of construction of unitary irreps for
different orbits of tensorial Poincar{\'e} is the construction of
field theories with corresponding symmetry algebra. For d=2+q
dimensional generalization of \ref{0} (connected with the SO(2,10)
hypothesis of \cite{Bars}) the corresponding field theories are
constructed for an orbit, which is a generalization of massless
particle orbit and for lowest representations of corresponding
little groups \cite{Man}\cite{Man2}. For lowest representations of
little groups of preonic orbits at dimensions 4, 6, ... the free
field theories, on the level of equations of motions, are actually
constructed in \cite{Vas}, with different motivation, coming from
higher spin theories in usual space-times. The first step towards
interaction in these theories is done for 2+2 dimensional theory
in \cite{mrl}.

In the next Section  Wigner's method is briefly described and
applied to the present case. Main results - the little groups for
massless case, more precisely, for the zero values of all
invariants (if those exist) are combined into Table 1. The
additional information on little groups with non-zero invariants
is combined into Table 2.

\section{Little groups of preons}

Modern supersymmetry algebras can be represented in a form

\begin{eqnarray}\label{6}
\left \{\bar{Q}^{j\beta},Q_{i\alpha} \right \} &=&
Z^{\;\;j\beta}_{i\alpha}
\end{eqnarray}
where supercharges $Q_{i\alpha}$ are subject to some constraints -
Majorana,Weyl, Symplectic-Majorana etc., r.h.s. matrix $Z$ is the
most generic matrix, that satisfies the same constraints. We shall
consider the minimal algebras, i.e. spinors $Q_{i\alpha}$ will be
the minimal possible in a given dimension, hence index i of the
group of R-invariance either will be absent, or will be that of
$SU(2)$ doublet, in the case of symplectic spinors. Matrix $Z$ can
be expanded over gamma-matrices basis and in that form r.h.s of
(\ref{6}) will become a usual combination of momentum $P_\mu$ and
"central charges" tensors $Z_{\mu\nu...}$. The bosonic subalgebra
is a semidirect product of Lorentz and R-symmetry group, from one
side, and Abelian group of momentum and central charges tensors,
from the other, so for construction of unitary irreps of that
algebra Wigner's method \cite{Wig}\cite{BR} can be applied. The
construction includes the following steps: classification of
orbits of action of Lorentz $\times $ R group on the numeric
tensor $Z$, determination of a stabilizer (little group) of some
(arbitrarily picked) point on the orbit, choice of any unitary
irrep of little group and finally induction on the whole tensorial
Poincar{\'e} group. The present paper is devoted to the discussion
of preonic orbits, i.e. the orbits with rank one Z matrix. Such
Z's have a form $Z=\lambda\lambda$, where $\lambda$ is a numeric
spinor, and stabilizer of Z is that of $\lambda$ plus
transformations which change the sign of $\lambda$, (which is
actually the property of full 360 degree rotation of spinor).
Also, such configuration of $Z$ is providing the maximum number of
surviving supersymmetries \cite {Band}\cite {Band2}. So, we come
to the problem of determination of orbits of action of Lorentz
group on the space of spinors $\lambda$, and calculation of
corresponding little groups.

The main results on little groups of preonic orbits are combined
into Tables 1 and 2. For dimensions $d=2\div 11$, with one time
dimensions, we list the type  of minimal spinor in corresponding
dimension (i.e. (pseudo)Majorana=(P)M, (pseudo)Symplectic
Majorana=PSM, Weyl=W, etc.), I - the number of independent
invariants, which can be constructed from corresponding spinor,
and the little group for the maximally non-compact case, more
exactly, for the zero values of all invariants. The necessary
additional information is the embedding of corresponding little
group into the group of symmetry, i.e. the product of Lorentz and
R-symmetry group (SO(3), if any). These embeddings are described
below, as well as explicit expressions for invariants, except d=7
and d=11 cases, where we don't know the complete minimal set of
independent invariants. Nevertheless, in the last case we can
claim, on the basis of some additional calculations, that these
invariants are actually invariant w.r.t. the d=12=2+10 Lorentz
group, which is well-known to be the automorphisms group of d=11
superalgebra. For obtaining all these results we use a combination
of few methods: sometimes we calculate an algebra of stabilizer
(little group) around an (arbitrary) given spinor, and calculate a
dimensionality of orbit at that point by the evident formula dim
(orbit)=(dim G) - dim (little group), where G is the product of
Lorentz and R-symmetry groups. From this we can find the number of
independent invariants, which can be constructed from a given
spinor, obviously that is a codimension of an orbit in a spinor
space, we denote that $I$ in the Tables. In few cases an algebraic
and group-theoretic considerations are applied, particularly when
considering the dimensional reductions of spinors. We use a
"mostly plus" metric, other notations and definitions of spinors
and corresponding matrices are based on \cite{Kugo}

Table 1. Massless preons' little groups.
\begin{equation}\label{aa}
\begin{array}{*{20}c}
   d & \begin{array}{l}
 {\rm{Spinor}} \\
 {\rm{type}} \\
 \end{array} &  \begin{array}{l}
 {\rm{I - invariants'}} \\
 {\rm{number}} \\
 \end{array} & \begin{array}{l}
 {\rm{Little}} \\
 {\rm{Group}} \\
 \end{array}   \\
   2 & {{\rm{MW}}} &  {\rm{0}} & {1}  \\
   3 & {\rm{M}} & {\rm{0}} & {T_1}   \\
   4 & {\rm{M}} &  {\rm{0}} & {T_2}   \\
   5 & {{\rm{PSM}}} &  {\rm{1}} & {SO(3)\ltimes T_3}   \\
   6 & {{\rm{SMW}}} &  {\rm{0}} & {SO(4)\ltimes T_4}   \\
   7 & {{\rm{SM}}} &  {\rm{3}} & {SO(4)\ltimes T_5}   \\
   8 & {{\rm{PM}}} &  {\rm{2}} & {SU(3)\ltimes T_6}   \\
   9 & {{\rm{PM}}} &  {\rm{1}} & {G_2\ltimes T_7}   \\
   {10} & {{\rm{MW}}} &  {\rm{0}} & {SO(7)\ltimes T_8}   \\
   {11} & {{\rm{M}}} &  {\rm{7}} & {SO(7)\ltimes T_9} &  \\
\end{array}
\end{equation}

Table 2. Massive preons' little groups.
\begin{equation}\label{aa}
\begin{array}{*{20}c}
   d & \begin{array}{l}
 {\rm{Spinor}} \\
 {\rm{type}} \\
 \end{array} &  \begin{array}{l}
 {\rm{I - invariants'}} \\
 {\rm{number}} \\
 \end{array} & \begin{array}{l}
 {\rm{Little}} \\
 {\rm{Group}} \\
 \end{array}   \\
   5 & {{\rm{PSM}}} &  {\rm{1}} & {SO(4)}  \\
   8 & {{\rm{PM}}} &  {\rm{2}} & {G_2}   \\
   9 & {{\rm{PM}}} &  {\rm{1}} & {SO(7)}  \\
\end{array}
\end{equation}
In the cases with $I=0$ the whole space of spinors (besides zero
point) is an orbit of corresponding Lorentz $\times$ R-invariance
group. Turning to the embeddings of little groups, we first
mention that they are not simply the subgroups of direct product
of Lorentz and R-symmetry groups (the later is trivial except the
SM(W) cases when it is SU(2)), but actually they should be the
subgroups of product of R-symmetry group with the little group of
particles - $SO(d-2)\ltimes T_{d-2}$ for massless and $SO(d-1)$
for massive cases. It is obvious from the fact that in all these
dimensions one can define the momenta $p^\mu$, corresponding to
$Z=\lambda\lambda$: $p_\mu \sim \bar{\lambda}\gamma^\mu\lambda$,
and since stabilizer of $Z$ evidently is stabilizer of $p_\mu$,
statement follows. Taking into account that Table 1 contains
little groups for the massless cases only (all invariants zero,
see below), we should describe embeddings of groups in the Table 1
into $R \times SO(d-2)\ltimes T_{d-2}$, which in turn has a
well-known embedding in $R\times SO(1,d-1)$. Additional remark is
that non-compact factors $T_{d-2}$ in Table 1 always coincide with
corresponding factor in $R \times SO(d-2)\ltimes T_{d-2}$, so we
need to describe embeddings of compact factors, only. Precise
embedding depends on a point on the orbit, we need to describe
embeddings up to equivalence, i.e. up to similarity transformation
in the group. The Table 2 contains results for massive cases,
which are also described below.

Cases d=3, 4 are evident, the absence of invariants means that the
whole space of spinors is one orbit, except the zero point. The
answer for the little group can be obtained by direct calculation,
for an arbitrary given spinor. The first non-trivial case is d=5,
with pseudo-symplectic-Majorana spinors. The massless particle
little group is SO(3), the same is R-symmetry, and SO(3) in the
Table 1 is just a diagonal subgroup of their direct product. The
only independent invariant is $m= i\lambda_\alpha^i \lambda_\beta
^j C^{\alpha\beta}\epsilon_{ij}$. Evidently, for any (non-zero)
values of this invariant little groups are the same. Existence of
this invariant means that the space of non-zero spinors is not a
single orbit, as at d=4, but an infinite set of orbits,
distinguished by the value of invariant. The above little group
corresponds to the zero value of that invariant. One can show,
that the mass of momenta $p^\mu=\lambda_\alpha^i \lambda_\sigma ^j
C^{\alpha\beta}\gamma_\beta^{\mu\sigma}\epsilon_{ij}$,
corresponding to $Z=\lambda\lambda$,  is exactly $m$:
$p_\mu^2=-m^2$. For an orbit with non-zero $m$ the little group
for preons is SO(4) (without any noncompact factor) and coincide
with little group of massive particle, the only additional detail
is that the embedding of this SO(4) into SO(3) of R-symmetry,
which is actually a set of two embeddings of two SO(3) factors of
SO(4) into SO(3), is non-zero, and depends on a particular point
on an orbit.

For d=6 the compact factor of massless little group is $SO(4)\sim
SO_L(3)\times SO_R(3)$ and $SO(4)$ group in the Table 1 is $\sim
SO_L(3)\times (SO_R(3)\times SO(3))_{diag}$ where last $SO(3)$ is
the group of R-symmetry.

At d=7 for the zero values of invariants (see discussion below for
d=11 case) the (compact part of the) little group is SO(4) which
is the same group described above for d=6, embedded in a natural
way (e.g. as a 4x4 sub-matrix in a left upper corner of 5x5
matrix, there is no other, non-equivalent representation) into
SO(5) group of massless particle at d=7.

At d=8 there are exactly two independent invariants: $m_1=i
\lambda_\alpha^i \lambda_\beta ^j C^{\alpha\beta}\epsilon_{ij}$
and $m_2=\lambda_\alpha^i \lambda_\sigma ^j
C^{\alpha\beta}\gamma_\beta^{9\sigma}\epsilon_{ij}$. The momenta
$p^\mu=\lambda_\alpha^i \lambda_\sigma ^j
C^{\alpha\beta}\gamma_\beta^{\mu\sigma}\epsilon_{ij}$ square into
$p_\mu^2=-(m_1^2+m_2^2)$, so for massless case we should have both
masses equal to zero. In that case the (compact part of the)
little group in Table 1 is $SU(3)$, and is a natural subgroup of
SU(4), which is isomorphic to the (compact part of the) massless
particle little group SO(6). At massive case, i.e. when one or two
of masses $m_1, m_2$ are non-zero, the little group is $G_2$. That
can be shown as follows. The little group should be a subgroup of
that of massive particle - SO(7). Acting by that group in a spinor
representation on the Weyl part of spinor $\lambda$, which is an
SO(7) spinor, we obtain the stabilizer $G_2$, since that is
exactly (one of the) definition(s) of $G_2$. It is easy to show,
that action of SO(7) on the anti-Weyl part of spinor $\lambda$
gives the same stabilizer, since Weyl and anti-Weyl parts of
$\lambda$ are connected by (pseudo)-Majorana condition.

At d=9 situation is similar to d=5: there is one invariant $m=i
\lambda_\alpha \lambda_\beta C^{\alpha\beta}$, the square of
momenta $p^\mu=\lambda_\alpha \lambda_\sigma
C^{\alpha\beta}\gamma_\beta^{\mu\sigma}$ is given by
$p_\mu^2=-m^2$. So the whole space of spinors is a set of orbits,
with different $m$, and there are two kinds of orbits: the first
one is massless, with $m=0$, the little group is $G_2\ltimes T_7$
and $G_2$ in its fundamental 7-dimensional representation is
embedded in SO(7), the compact factor of massless little group.
Second one is massive, with little group SO(7), naturally a
subgroup of the little group of massive particle, SO(8). It is
clear from the triality: the problem of finding the stabilizer of
a given d=1+8 PM spinor leads to the problem of finding a
stabilizer of spinor under SO(8) (the stabilizer of its momenta
$p_\mu$), which, in turn, due to the triality property is
equivalent to finding a stabilizer of eight-dimensional Euclidean
vector under an SO(8) rotations, which evidently is SO(7)
subgroup.

At d=10 little group is SO(7) \cite{mrl3}, embedded, in its spinor
representation, into a fundamental representation of SO(8)
\cite{Oct}.

Finally, at d=11 we have little group SO(7), embedded, as at d=10,
in its spinor representation into SO(8), which, in turn, naturally
is a subgroup of d=11 massless SO(9). We cannot present a complete
set of independent invariants (although, of course, a lot of them
can be easily constructed), but we can state, that all values of
those invariants are zero for an orbit of Table 1 (in our
representation of gamma-matrixes, that is an orbit of a spinor
with unites at first and sixth places, zero otherwise). That
follows from the fact that, from one hand, these invariants are
some homogeneous polynomials over $\lambda$, and, from the other
hand, we can find a Lorentz transformation, which simply rescales
a given spinor on the orbit, so the non-zero value of these
polynomials will be rescaled, which contradicts to their
invariance. This argument actually works in other dimensions too.
The interesting further remark on this d=11 case is the following.
It is well-known, that d=11 algebra (\ref{0}) can be represented
in d=2+10=12 dimensional form, since corresponding Lorentz group
SO(2,10) is an automorphisms group of (\ref{0}). In that form
momenta and second rank tensor combine into one d=12 second-rank
tensor, and fifth-rank tensor $Z_{\mu \nu \lambda \rho \sigma} $
is interpreted as 12d self-dual sixth rank tensor, 11d Majorana
supercharges become 12d Majorana-Weyl supercharges, and algebra
receives the form:
\begin{eqnarray}
\left\{ \bar{Q},Q\right\} &=&\Gamma ^{\mu \nu }P_{\mu \nu }+\Gamma
^{\mu \nu \lambda \rho \sigma \delta }Z_{\mu \nu \lambda \rho
\sigma \delta }^{+}
\label{2} \\
\mu \nu ,... &=&0^{\prime },0,1,...10  \nonumber
\end{eqnarray}
One can study orbits of the same spinor under the extended group
SO(2,10). The remarkable fact is that the orbit has the same
dimensionality, 25, as under the action of SO(1,10) group, so
coincide with 11d orbit. That follows from the fact, that although
corresponding little group has more complicated structure, but its
dimensionality is 41, so dimensionality of orbit is dimensionality
of Lorentz group (66) minus dimensionality of little group (41),
i.e. 25. So, assuming this coincidence for any spinor, we conclude
that the number of independent invariants is the same, 7, they are
SO(2,10) invariant, and can also serve as SO(1,10) invariants, so
they are exactly an 11d invariants, so we prove that 11d Majorana
spinors invariants can be chosen in 12d invariant form. In other
words, 7 invariants of 11d case of Table 1 can be written in a 12d
invariant form, which is non-trivial statement and provide some
support for ideas of 12d invariance of M-theory.

\section{Acknowledgements}

This work is supported partially by INTAS grant \#99-1-590. We are
indebted to R.Manvelyan for discussions. R.M. is indebted to
D.Sorokin for discussion and to A.Sagnotti for discussions and for
hospitality at "Tor Vergata" University.

\end{document}